\documentclass[12pt,fleqn]{article}
\usepackage{epsfig,times,latexsym,enumerate,lscape}
\newcommand{\prd}{Phys. Rev. D }
\newcommand{\pp}{Preprint gr-qc/}

\begin{document}

\title{Data analysis of continuous gravitational wave: Fourier transform-I}
\author{D.C. Srivastava$^{1,2}$\thanks{e-mail: dcsrivastava@now-india.com} 
      $\,$ and S.K. Sahay$^1$\thanks{e-mail: ssahay@iucaa.ernet.in}\thanks{Present address: Inter University Centre for Astronomy and Astrophysics, Post Bag 4, Ganeshkhind, Pune - 411007, India}\\ \\
\normalsize $^1$Department of Physics, DDU Gorakhpur University, Gorakhpur-273009, U.P., India.\\
\normalsize $^2$Visiting Associate, Inter University Centre for Astronomy and Astrophysics, Post  \\
\normalsize Bag 4, Ganeshkhind, Pune-411007, India.}
\date{}
\maketitle
\begin{abstract}
We present the Fourier Transform of a continuous gravitational wave.
We have analysed the
data set for one day observation time and our analysis is
applicable for arbitrary location of detector and source. We have taken
into account the effects arising due to rotational as well as orbital motions
of the earth.
\end{abstract}

\section{Introduction}
\indent The first generation of long-baseline laser interferometers 
and ultra cryogenic bar detectors will start collecting data very soon.
The network of detectors is expected not only to confirm the existence of
gravitational waves (GW) but will also provide
the information about the structure and the dynamics of its source. At the
present stage, the data analysis depends largely upon the study of the expected
characteristic of its potential sources and the waveforms. Majority of the
experimental searches is focused on the detection of burst 
and {\it chirp\/} signals. However, the interest in the data analysis for 
continuous gravitational wave (CGW) signals is growing. A prime example of sources of this type is
a spinning neutron star. Many research groups around the globe are working
extensively on the data analysis for spinning neutron stars
(Jaranwoski, et.al 1998, 1999, 2000; Brady et. al. 1998, 2000; 
Kr\'olak 1999). \\

\par The detection of GW signals in the noisy output of the detectors has its
own problems, not the least of which is the sheer volume of data analysis. Bar 
detectors have essentially the same problems as interferometers 
in reference to CGW sources. Each detector produces a single data stream that may contain
many kinds of signals. Detectors don't point, but rather sweep their broad 
quadrupolar beam pattern across the sky as the earth moves. So possible 
sources could be anywhere on the sky and accordingly the data analysis
algorithms need to accommodate signals from any arbitrary location of its
source.\\

\par In this and the subsequent paper we present analysis of Fourier Transform (FT) of the output
data of a ground based laser interferometer. The output data has broad band
noise and the signal is to be extracted out of it. For
this, one has to enhance signal-to-noise ratio (SNR). This is achieved by analyzing long
observation time data as SNR is directly proportional to the square root of
observation time ($\sqrt {T_o}$).
However, in a data for long duration, the monochromatic signal gets Doppler
modulated due to (i) the orbital motions of the earth around the sun and (ii) the spin of 
the earth. The frequency modulation (FM) will spread the signal in a very large number
of bins depending on the source location and the frequency. In addition to
this, there is amplitude modulation (AM). As we will see in the sequel the
amplitude of the detector output consists of simple harmonic terms with
frequencies $w_{rot}$ and $2 w_{rot}$ where, $w_{rot}$ stands for angular
rotational frequency of the earth. Accordingly, the
AM results in splitting of the FT into frequencies $\pm w_{rot}$
and $\pm 2 w_{rot}$.\\

\par In the next section we present the noise free response of the laser
interferometric detector and obtain the explicit beam pattern functions. In
section 3 we discuss the Doppler effect and obtain Fourier transform (FT) of the FM signal for
arbitrary source and detector locations taking into account the earth's
rotational motion about its axis and its revolution around the sun. In
section 4 the FT of the Doppler modulated complete response
of the detector is obtained. Section 5 is the conclusion of the
paper.

\section{The noise free response of detector: Beam pattern and Amplitude
modulation}
\label{sec:nfr}
Let a plane GW fall on a laser interferometer and produce changes in the
arms of the detector. In order to express these changes quantitatively we have 
to specify the wave and the detector. Let $X Y Z$ and $x y z$ represent
respective frames characterising the wave and the detector. We assume the direction
of propagation of the wave to be the $Z$ axis and the vertical at the place of
the detector to be the $z$ axis. The difference of the changes $\delta l$ in the
arm lengths of the detector may be given via

\begin{equation}
\label{eq:deltal}
R(t) = \frac{\delta l}{l_o} = - \sin 2\Omega\left[\left( {\bf A}^x_X{\bf A}^y_X -
{\bf A}^x_Y{\bf A}^y_Y\right)h_+ + 
 \left( {\bf A}^x_X{\bf A}^y_Y + {\bf A}^x_Y{\bf A}^y_X\right) h_\times\right]
\end{equation}

\noindent where $l_o$ is the normal length of the arms of the detector and $2\Omega$
expresses the angle between them (Schutz and Tinto, 1987). The matrix $\left( {\bf A}^j_K \right)$
represents the transformation expressing the rotations to bring the wave
frame $(X , Y , Z)$ to the detector frame $(x , y , z)$. The direction of
the source may be expressed in any of the coordinates
employed in Spherical Astronomy. However, we find it convenient to define it
in Solar System Barycentre (SSB) frame, $(X' , Y' , Z')$. This SSB frame is
nothing but
astronomer's ecliptic coordinate system. Let $\theta$ and $\phi$ denote the celestial
colatitude and celestial longitude of the source. These coordinates are related
to the right ascension, $\bar{\alpha}$ and the declination, $\bar{\delta}$ of
the source via

\begin{equation}
\left.\begin{array} {rcl}
\vspace{0.2cm}
\cos\theta & = & \sin\bar{\delta}\cos\epsilon - \cos\bar{\delta}\sin
\epsilon\sin\bar{\alpha}\\
\vspace{0.2cm}
\sin\theta\cos\phi & = & \cos\bar{\delta}\cos\bar{\alpha} \\
\sin\theta\sin\phi & = & \sin\bar{\delta}\sin\epsilon + \cos\bar{\delta}
\cos\epsilon\sin\bar{\alpha}
\end{array}\right\}
\vspace{0.3cm}
\end{equation}

\noindent where $\epsilon$ represents obliquity of the ecliptic. We choose $x$ axis as the bisector of the angle between the arms of the
detector. At this stage the orientation of the detector in the horizontal plane is arbitrary.
It is assigned with the help of the angle $\gamma$ which the $x$ axis makes with
the local meridian. The location of the detector on the earth is characterised by
the angles, $\alpha$ - colatitude and $\beta$ - the local sidereal time, expressed
in radians.
The transformation matrix $\left( {\bf A}^j_K\right)$ may be expressed as

\begin{equation}
{\bf A\/} = {\bf DCB \/}
\end{equation}

\noindent where

\noindent ${\bf B\/}:$ rotation required to bring $X Y Z$ to $X' Y' Z'$\\
${\bf C\/}:$ rotation required to bring $X' Y' Z'$ to $x' y' z'$\\
${\bf D\/}:$ rotation required to bring $x' y' z'$ to $x y z$\\

\noindent Here $x' y' z'$ represents the frame associated
with the earth. The Euler angles
defining the corresponding rotation matrices are given via

\begin{equation}
\left.\begin{array}{ccl}
\vspace{0.2cm}
{\bf B \/}& : & (\theta , \phi , \psi )\\
\vspace{0.2cm}
{\bf C \/}& : & (0 , \epsilon , 0 )\\
{\bf D \/}& : & (\alpha , \beta + \pi /2 , \gamma - \pi /2 )
\end{array}\right\}
\end{equation}

\noindent where $\psi$ is a measure of the polarisation of the wave (Goldstein, 1980). \\
Let us write Eq.~(\ref{eq:deltal}) as

\begin{equation}
\label{eq:rt}
R(t) = \frac{\delta l}{l_o} = - \sin 2\Omega \left[ F_+h_+ + F_\times h_\times \right]
\end{equation}

The functions $F_+$ and $F_\times$ involve the angles $\theta , \phi , \psi ,
\epsilon  , \alpha , \beta , \gamma$ and express the effect of the interaction of
the wave and the detector. These are called antenna or beam patterns. The explicit 
functional dependences of $F_+$ and $F_{\times}$ are given by Jotania and Dhurandhar (1994). It is pointed out that there are few minor errors in these expressions and these are later corrected by the authors in Jotania (1994). 
These functions appear complicated but may be written in simpler form by
introducing following abbreviations.

\begin{equation}
\left.\begin{array}{lcl}
\vspace{0.2cm}
U &=& \cos \alpha \cos \beta \cos \gamma - \sin \beta \sin \gamma\, , \\
\vspace{0.2cm}
V &= & - \cos \alpha \cos \beta \sin \gamma - \sin \beta \cos \gamma \, ,\\
\vspace{0.2cm}
X &= &\cos \alpha \sin \beta \cos \gamma + \cos \beta \sin\gamma \, ,\\
\vspace{0.2cm}
Y &=& - \cos \alpha \sin \beta \sin \gamma + \cos \beta\cos \gamma \, 
\end{array} \right\}\hspace{4.05cm}
\end{equation}

\vspace{0.2in}

\begin{equation}
\left.\begin{array}{lcl}
\vspace{0.2cm}
L &= &\cos\psi\cos\phi - \cos\theta\sin\phi\sin\psi\, ,\\
\vspace{0.2cm}
M &=& \cos\psi\sin\phi + cos\theta\cos\phi\sin\psi\, , \\
\vspace{0.2cm}
N &=& - \sin\psi\cos\phi - cos\theta\sin\phi\cos\psi\, ,\\
\vspace{0.2cm}
P& = &-\sin\psi\sin\phi + cos\theta\cos\phi\cos\psi \, ,\\
Q &=&\sin\theta\sin\phi\, ,\qquad R\; =\; \sin\theta\cos\phi \, ,
\end{array} \right\} \hspace{4.0cm}
\end{equation}

\vspace{0.2in}

\begin{equation}
\left.\begin{array}{lcl}
\vspace{0.2cm}
A &= &2 X Y\cos^2\epsilon - \sin^2\epsilon \sin^2\alpha \sin 2\gamma +
 \sin 2\epsilon ( X \sin \alpha \sin \gamma - Y \sin \alpha \cos\gamma ) \, ,\\
\vspace{0.2cm}
B& = & 2 X Y\sin^2\epsilon - \cos^2\epsilon \sin^2\alpha \sin 2\gamma -
\sin 2\epsilon ( X \sin \alpha \sin \gamma - Y \sin \alpha \cos \gamma ) \, ,\\
\vspace{0.2cm}
C& =& \cos\epsilon ( Y U + X V ) + \sin\epsilon ( U \sin \alpha \sin
\gamma  - V\sin\alpha \cos \gamma )\, ,\\
\vspace{0.2cm}
D &= &- \sin\epsilon ( Y U + X V ) + \cos\epsilon ( U \sin \alpha
\sin\gamma - V \sin\alpha \cos \gamma ]\, ,\\
\vspace{0.2cm}
E& = &- 2 X Y \cos\epsilon \sin\epsilon - \cos\epsilon\sin\epsilon
\sin^2\alpha\sin 2\gamma + \cos 2\epsilon ( X \sin\alpha\sin\gamma - Y \sin \alpha \cos\gamma )
\end{array} \right\}
\end{equation}

\vspace{0.2in}

\noindent After straight-forward substitutions one obtains

\begin{eqnarray} F_+(t) & = &{1\over 2}\left[ 2 ( L^2 - M^2 ) U V + (
N^2 - P^2 ) A \: +  ( Q^2 - R^2) B \right] + ( L N - M P ) C
\nonumber\\
\label{eq:fplust}
 &&+\, ( L Q + M R ) D + ( N Q + P R ) E\, ,\\
F_\times (t)&
= & 2 L M U V + N P A - {1\over 2} B \sin^2\theta\sin 2\phi\: +
(LP + M N ) C \nonumber\\
 &&+\, ( M Q - L R ) D + ( P Q - N R ) E
\label{eq:fcrosst}
\end{eqnarray}

\vspace{0.2in}

\noindent The compactification obtained here arises due to the fact that above
introduced abbreviations find places in the transformation matrices as follows

\vspace{0.2in}

\begin{equation}
{\bf B \/}= \left( \begin{array}{ccc}
L &   N & Q \\
M & P & - R\\
\sin\theta\sin\psi & \sin\theta\cos\psi & \cos\theta \end{array}\right) 
\end{equation}

\vspace{0.2in}

\begin{equation}
{\bf C \/}= \left( \begin{array}{ccc}
1 &   0 & 0 \\
0 & \cos\epsilon & \sin\epsilon\\
0 & - \sin\epsilon & \cos\epsilon\end{array}\right) 
\end{equation}

\vspace{0.2in}

\begin{equation}
{\bf D \/}= \left( \begin{array}{ccc}
U &   V & \sin\alpha\cos\beta \\
X & Y & \sin\alpha\sin\beta\\
- \sin\alpha\cos\gamma & \sin\alpha\sin\gamma & \cos\alpha \end{array}\right)\; ;
\end{equation}

\vspace{0.2in}

\noindent After algebraic manipulation Eqs.~(\ref{eq:fplust})
and~(\ref{eq:fcrosst}) may be expressed as

\begin{eqnarray}
\label{eq:fps}
F_+(t) &= &F_{1_+}\cos 2\beta + F_{2_+}\sin 2\beta + F_{3_+}\cos \beta + 
 F_{4_+}\sin \beta + F_{5_+}\; ;
\end{eqnarray}
\begin{eqnarray}
\label{eq:fcs}
F_\times (t)&= &F_{1_\times}\cos 2\beta + F_{2_\times}\sin 2\beta +
F_{3_\times}\cos \beta + 
 F_{4_\times}\sin \beta +
F_{5_\times}\end{eqnarray}

\vspace{0.2in}

\noindent where $F_{i _+}$ and $F_{i_\times}$ $( i = 1,2,3,4,5 )$ are time
independent expressions given via

\begin{equation}
\left.\begin{array}{lcl}
\vspace{0.2cm}
F_{1_+} &= & - 2 G \cos\alpha\cos 2\gamma + {H \sin 2\gamma\over 2}(
\cos^2\alpha + 1 ) \, ,\\
\vspace{0.2cm}
F_{2_+} &= &H \cos\alpha\cos 2\gamma + G\sin 2 \gamma (\cos^2\alpha + 1 )\, ,\\
\vspace{0.2cm}
F_{3_+} &= &I\sin\alpha\cos 2\gamma + J \sin 2\alpha\sin 2\gamma\, ,\\
\vspace{0.2cm}
F_{4_+} &= &2 J \sin\alpha\cos 2\gamma - {I\over 2}\sin 2\alpha\sin 2\gamma\, ,\\
F_{5_+} &=& {3\sin^2\alpha\sin 2\gamma\over 2}[ H + L^2 - M^2 ]\, ,
\end{array} \right\} \; ;
\end{equation}

\vspace{0.2in}

\begin{equation}
\left.\begin{array}{lcl}
\vspace{0.2cm}
G& =& {1\over 2}[ ( L Q + M R )\sin\epsilon - ( L N - M P )\cos\epsilon ]\, ,\\ 
\vspace{0.2cm}
H &=& {1\over 2}[ ( N^2 - P^2 ) \cos^2\epsilon
- ( L^2 - M^2 ) + 
 ( Q^2 - R^2 )\sin^2\epsilon - ( N Q + P R )\sin 2\epsilon ]\, ,\\
\vspace{0.2cm}
I& = &{1\over 2}[ (Q^2 - R^2 ) \sin 2\epsilon - (
N^2 - P^2 )\sin 2\epsilon - 2 ( N Q + P R )\cos 2\epsilon ]\, ,\\
J& =& {1\over 2}[ ( L N - M P )\sin\epsilon + ( L Q + M R
)\cos\epsilon ] \\
\end{array} \right\}
\end{equation}

\vspace{0.2in}

\noindent Let us note that $ F_{i_\times} $ is related to $ F_{i_+} $ via

\begin{eqnarray}
F_{i_\times}( \theta , \phi , \psi , \alpha , \beta , \gamma ,
\epsilon )& = & F_{i_+}( \theta , \phi - {\pi\over 4}, \psi , \alpha ,
\beta , \gamma , \epsilon ) \; ; \qquad i\; = \;1,2,3,4,5 
\end{eqnarray}

\noindent This symmetry is representative of the quadrupolar nature of the
detector and the wave. The detectors at different orientations will record
different amplitudes in their responses. The explicit beam pattern functions may
be computed easily for any instant of time. Due to the
symmetries involved in $F_+$ and $F_\times$ it is sufficient to
evaluate either of the beam patterns.\\

\par  The amplitude modulation of the received signal is a
direct consequence of the non-uniformity of the sensitivity pattern. As
remarked earlier they are fairly complicated function
of their arguments. Equations~(\ref{eq:fps}) and~(\ref{eq:fcs})
reveal that the monochromatic signal
frequency will split, due to AM, into five frequencies. This results in the distribution of energy in various
frequencies and consequent reduction of the amplitude of the signal. The periodicity of the beam
pattern $F_+$ and $F_\times$ with a period equal to one sidereal day
 is due
to the diurnal motion of the Earth.

\section{Doppler shift and Frequency modulation}
\label{sec:fmd}
\indent The frequency of a monochromatic signal will be Doppler shifted
due to the translatory motion of the detector acquired from the motions of
the earth. Let us  consider a CGW signal of constant frequency $f_o$.
The frequency $f'$ received at the instant $t$ by the detector is given by

\begin{equation}f'(t) = f_o\gamma_o\left( 1 + {{\bf v . n} \over c}(t)\right)\quad ; \quad
\gamma_o = \left(1 - {v^2\over c^2}\right)^{- 1/2}
\end{equation}

\noindent  where ${ \bf\textstyle n}$ is the unit vector from the antenna to the source,
${\displaystyle {\bf v}}$ is the relative velocity of the source and the antenna,
and $c$ is the velocity of light. The unit vector ${ \bf\textstyle n}$ from the
antenna to the source, in view of the  fact that the distance of the source is 
very large compared to the average distance of the centre of the SSB frame and the detector, may
be taken parallel to the unit vector drawn from the centre of the SSB frame to
the source. Hence,

\begin{equation}
{ \bf\textstyle n}\quad =\quad \left(\sin\theta\cos\phi \quad , \quad \sin\theta\sin\phi
\quad , \quad \cos\theta\right)
\end{equation}

\noindent As {\bf v\/} keeps on changing continuously
both in its amplitude and direction $f'$ is a continuous function of t. Further,
since $v \ll c $ we take $\gamma_o = 1$. \\

\noindent The radius vector ${\displaystyle\bf r\/}(t)$ in
the SSB frame is given by (Kanti, et al. 1996)

\begin{eqnarray}
{\bf r\/}(t) & = & \left[ R_{se} \cos (w_{orb} t) + R_e \sin\alpha \cos\beta\right.\, , \;
R_{se} \sin (w_{orb} t) + R_e \sin\alpha\sin\beta\cos\epsilon - \nonumber \\
&& \left. R_e \cos\alpha \sin\epsilon \, ,  
 R_e \sin\alpha \sin\beta\sin\epsilon + R_e \cos\alpha \cos\epsilon\right]\; ;
\end{eqnarray}

\begin{equation}
\label{eq:beta}
\beta = \beta_o + w_{rot}t
\end{equation}

\noindent  where $R_{e}$, $R_{se}$ and $w_{orb}$ represent respectively the
earth's radius, average distance between earth's centre from the origin of
SSB frame and the orbital angular velocity of the earth. Here $t$ represents the
time in seconds elapsed from the instant the sun is at the Vernal Equinox and $\beta_o$
is the local sidereal time at that instant. The Doppler shift is now given via 

\begin{eqnarray}
\frac{f' - f_o}{f_o}&=&{{\bf v . n}\over c}(t)\; = \; {\Huge{{\bf\dot{r}_t . n}\over c}} \; = \;
 { R_{se} w_{orb}\over c}\sin\theta\sin ( \phi - w_{orb} t ) + \nonumber\\ 
&& {R_e w_{rot} \over c} \sin\alpha \left[ \sin\theta\{ \cos\beta\cos\epsilon\sin\phi  - 
 \cos\phi\sin\beta \}  + \cos\beta\sin\epsilon\cos\theta \right]
\label{eq:ds}
\end{eqnarray}

\noindent The phase $\Phi (t)$ of the received signal is given by

\begin{eqnarray}
 \Phi (t) & = & 2 \pi \int_0^t f' (t') dt'  \nonumber\\
& = & 2 \pi f_o \int_0^t  \left[ 1 + {{\bf v . n}\over c}
(t') \right]dt' 
\end{eqnarray}

\noindent Here we assume the initial phase of the wave to be zero.
\noindent After straight-forward calculation we obtain

\begin{eqnarray}
\Phi (t) & = &  2\pi f_o \left[ t + {R_{se}\over c} \sin
\theta\cos\phi' + \right.
{R_e\over c}\sin\alpha \{\sin\theta (\sin\beta\cos\epsilon\sin\phi + \cos\phi\cos\beta) + \nonumber \\
&& \sin\beta\sin\epsilon\cos\theta\} -  {R_{se}\over c}\sin\theta\cos\phi - 
 {R_e\over c}\sin\alpha\{\sin\theta (\sin\beta_o\cos\epsilon\sin\phi 
 +  \nonumber \\ 
&& \cos\phi\cos\beta_o) + 
 \left. \sin\beta_o\sin\epsilon\cos\theta\} \right]\nonumber \\
& =& 2\pi f_o t + {\cal Z}\cos (w_{orb} t - 
\phi ) + {\cal P}\sin (w_{rot}t) + {\cal Q}\cos (w_{rot}t) - {\cal R} - {\cal Q} \nonumber\\
& = & 2\pi f_o t + {\cal Z}\cos (a\xi_{rot} - \phi ) + 
{\cal N}\cos (\xi_{rot} - \delta ) - {\cal R} - {\cal Q}
\label{eq:phase}
\end{eqnarray}

\noindent where \\
\begin{equation}
\label{eq:daypq}
\left.\begin{array}{lcl}
\vspace{0.2cm}
{\cal P}& = & 2\pi f_o {R_e\over c} \sin\alpha (\cos\beta _o(\sin
\theta \cos\epsilon \sin\phi + \cos\theta \sin\epsilon )
 - \sin\beta _o \sin\theta \cos\phi )\, ,\\
\vspace{0.2cm}
{\cal Q}& = & 2\pi f_o {R_e\over c}\sin\alpha (\sin\beta _o (\sin\theta \cos
\epsilon \sin\phi + \cos\theta \sin\epsilon )
+ \cos\beta _o \sin\theta\cos\phi ) \, ,\\
\vspace{0.2cm}
{\cal N}& = & \sqrt{ {\cal P}^2 + {\cal Q}^2 }\, ,\\
\vspace{0.2cm}
{\cal Z}& = & 2\pi f_o {R_{se}\over c}\sin\theta \, , \\
{\cal R}& = & {\cal Z}\cos\phi\, ,\\
\end{array} \right\}
\end{equation}

\vspace{0.2in}

\begin{equation}
\label{eq:daydelta}
\left.\begin{array}{lcl}
\vspace{0.2cm}
\delta & = &  \tan^{- 1}\frac{{\cal P}}{{\cal Q}}\, ,\\
\vspace{0.2cm}
\phi' & = & w_{orb}t - \phi\, ,\\
\vspace{0.2cm}
\xi_{orb} & = & w_{orb}t\; = \; a\xi_{rot};\quad  a \;= \; w_{orb}/w_{rot}\; \approx \; 1/365.26\, , \\
\xi_{rot} & = & w_{rot}t 
\end{array} \right\}
\end{equation}

\noindent The two polarisation states of the signal can be taken as

\begin{equation}
h_+(t) = h_{o_+}\cos [\Phi (t)] 
\label{eq:hpt}
\end{equation}

\begin{equation}
h_\times (t) = h_{o_\times}\sin [\Phi (t)]
\label{eq:hct}
\end{equation}

\noindent $h_{o_+}$, $h_{o_\times}$ are the time independent amplitude of
$h_+(t)$, and  $ h_\times (t)$ respectively.\\ 

\indent To understand the nature of the FM let us consider the function 

\begin{equation}
h(t) = \cos[\Phi (t)]
\label{eq:cosphit}
\end{equation}

\noindent and analyse it for one day observation data. The FT is given via

\begin{equation}
\left[\tilde{h}(f)\right]_d = \int_0^T \cos[\Phi (t)]e^{-i2\pi ft}dt\; ; \qquad
 T =\; one \;sidereal \; day \; = 86164 \; sec.
\label{eq:hf1}
\end{equation}

\noindent This splits into two terms as 

 \begin{equation} \left[\tilde{h}(f)\right]_d = I_{\nu_-} + I_{\nu_+} \; ;
 \label{eq:inu}
 \end{equation}

 \begin{eqnarray}
\label{eq:dayinu}
 I_{\nu_-}& =& {1\over 2 w_{rot}}\int_0^{2\pi} e^{i \left[
 \xi\nu_- +
{\cal Z}\cos (a\xi - \phi ) + {\cal N}\cos (\xi - \delta ) -{\cal R}  - {\cal 
Q} \right] } d\xi\, ,\\
I_{\nu_+}& = &  {1\over 2 w_{rot}}\int_0^{2\pi} e^{- i \left[
\xi\nu_+
+ {\cal Z} \cos (a\xi - \phi ) + {\cal N}\cos (\xi - \delta ) - {\cal R} - {
\cal Q} \right] } d\xi\, , \\
\nu_{\mp}& = & \frac{f_o \mp f}{f_{rot}} ; \quad \xi \; = \; \xi_{rot} \; = \;
w_{rot}t 
\end{eqnarray}

\noindent Numerical result shows that $I_{\nu_+}$ oscillates very fast and
contributes very little to $\left[\tilde{h}(f)\right]_d$. Hence, hereafter, we drop $I_{\nu_+}$ 
from Eq.~(\ref{eq:inu}) and write $\nu$ in place of $\nu_-$. Using 
the identity

\begin{equation}
e^{\pm i\kappa\cos\vartheta} = J_o (\pm\kappa) + 2 \sum_{l = 1}^{l = 
\infty} i^l J_l (\pm\kappa)\cos l\vartheta
\label{eq:bessel}
\end{equation}

\noindent we obtain
\begin{eqnarray}
\left[\tilde{h}(f)\right]_d &\simeq & \frac{1}{2 w_{rot}} e^{i ( 
- {\cal R} - {\cal Q})} \int_0^{2\pi} e^{i\nu\xi} \left[ J_o( {\cal Z} ) + 2 
\sum_{k = 1}^{k =  \infty} J_k ({\cal Z}) i^k \cos k (a\xi - \phi )\right] 
 \nonumber \\
&& \times\,\left[J_o( {\cal N} ) + 2 \sum_{m = 1}^{m =  \infty} J_m ({\cal N}) i^m 
\cos m (\xi - \delta )\right] d\xi
\end{eqnarray}

\noindent where $J$ stands for the Bessel function of the first kind. After 
performing the integration we get

\begin{equation}
\label{eq:hfd}
\left[\tilde{h}(f)\right]_d  \simeq  \frac{\nu}{2 w_{rot}} \sum_{k  =  - 
\infty}^{k = \infty} \sum_{m = - \infty}^{m =  \infty} e^{ i {\cal A}}{\cal 
B}[ {\cal C} - i{\cal D} ] \; ; \;
\end{equation}  

\vspace{0.2in}

\begin{equation}
\left.\begin{array}{lcl}
\vspace{0.2cm}
{\cal A}&  = &{(k + m)\pi\over 2} - {\cal R} - {\cal Q} \nonumber \\
\vspace{0.2cm}
{\cal B} & = & {J_k({\cal Z}) J_m({\cal N})\over {\nu^2 - (a k + m)^2}} \nonumber \\
\vspace{0.2cm}
{\cal C} &= & \sin 2\nu\pi \cos ( 2 a k \pi - k \phi - m \delta ) - 
{ a k + m \over \nu}\{\cos 2 \nu \pi \sin ( 2 a k \pi - k \phi - m \delta )
+ \nonumber \\ 
&& \sin ( k \phi + m \delta )\}\nonumber \\
\vspace{0.2cm}
 {\cal D} & = & \cos 2\nu\pi \cos ( 2 a k \pi - k \phi - m \delta ) + 
 {k a +m \over \nu}\sin 2 \nu \pi \sin ( 2 a k \pi - k \phi - m \delta )
 - \nonumber \\ 
&&\cos ( k \phi + m \delta )  \nonumber 
\end{array} \right\}
\end{equation}

\vspace{0.2in}

\noindent The FT of the two polarisation states of the wave can now be written as

\begin{eqnarray}
\label{eq:hfpd}
\left[\tilde{h}_+(f)\right]_d&=&h_{o_+}\left[\tilde{h}(f)\right]_d \nonumber \\
&\simeq & \frac{\nu h_{o_+}}{2 w_{rot}} \sum_{k  =  - \infty}^{k =  
\infty} \sum_{m = - \infty}^{m =  \infty} e^{ i {\cal A}}{\cal B}[ {\cal C} - i{\cal D} ] \; ;
\end{eqnarray}

\begin{eqnarray}
\label{eq:hfcd}
\left[\tilde{h}_\times (f)\right]_d &=&- i h_{o_\times}\left[\tilde{h}(f)\right]_d \nonumber \\
&\simeq & \frac{\nu h_{o_\times}}{2 w_{rot}} \sum_{k  =  - \infty}^{k =  
\infty} \sum_{m = - \infty}^{m =  \infty} e^{ i {\cal A}}{\cal B}[ {\cal D} - i{\cal C} ] 
\end{eqnarray}

\begin{figure}
\centering
\epsfig{file=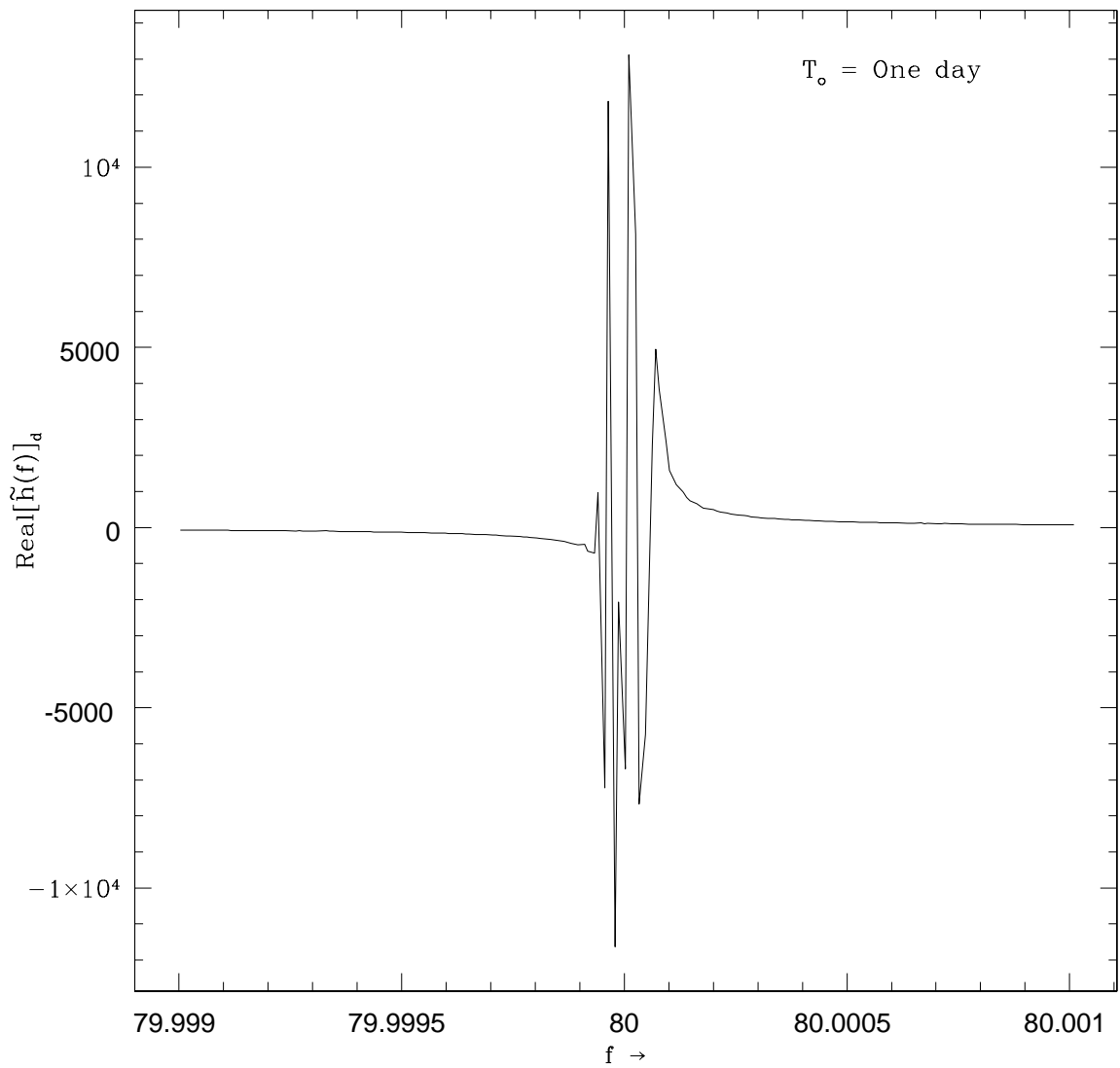,height=16.0cm}
\caption{FT of a FM signal frequency, $f_o = 80$ Hz, from a source located at 
$(\pi /36 , \pi )$ with a resolution of $1.16 \times 10^{-5}$ Hz}
\label{fig:drealfmrt}
\end{figure}

\begin{figure}
\centering
\epsfig{file=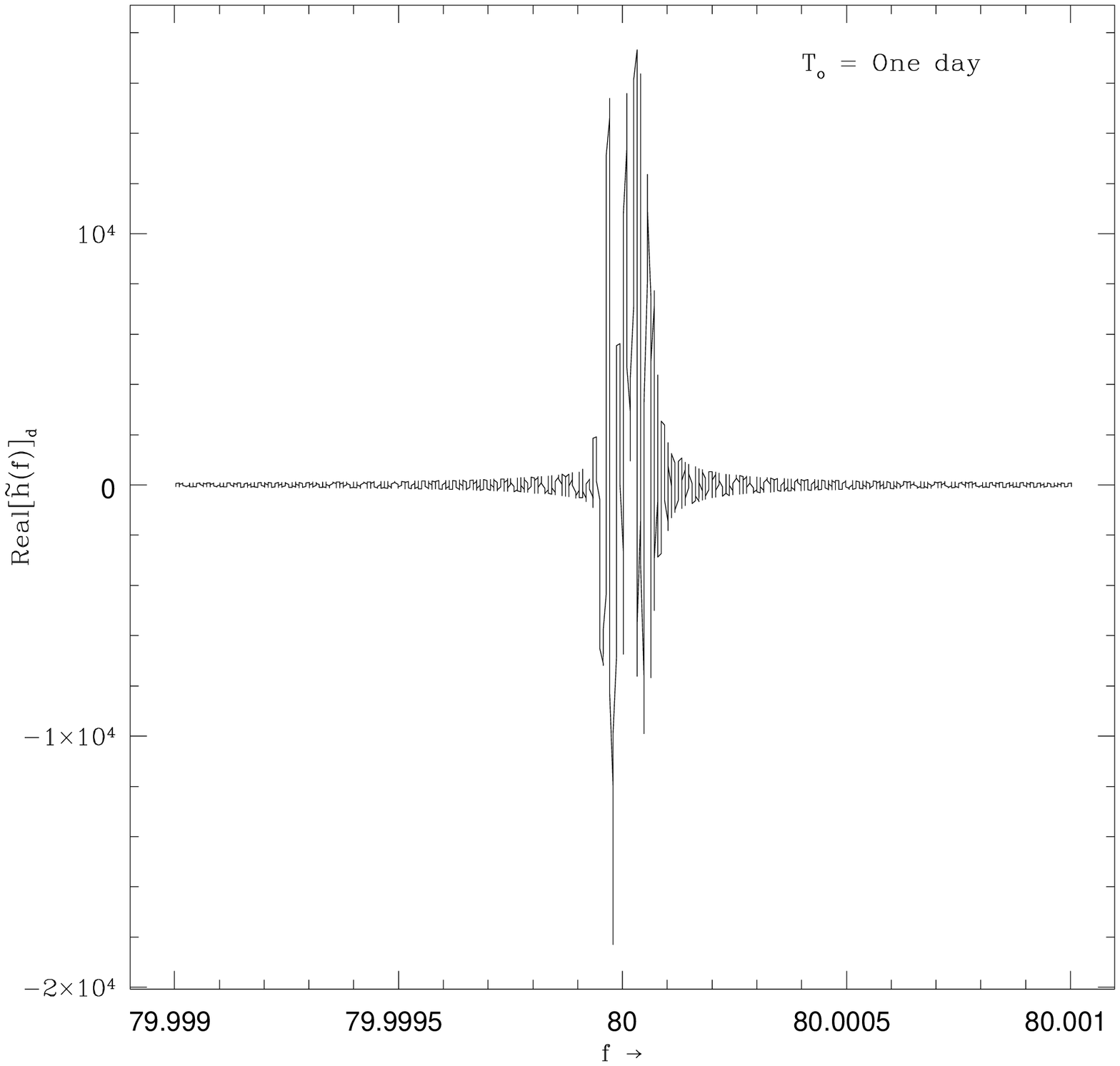,height=16.0cm}
\caption{FT of a FM signal frequency, $f_o = 80$ Hz, from a source located at 
$(\pi /36 , \pi )$ with a resolution of $10^{-6}$ Hz}
\label{fig:drealfmr6}
\end{figure}

\begin{figure}
\centering
\epsfig{file=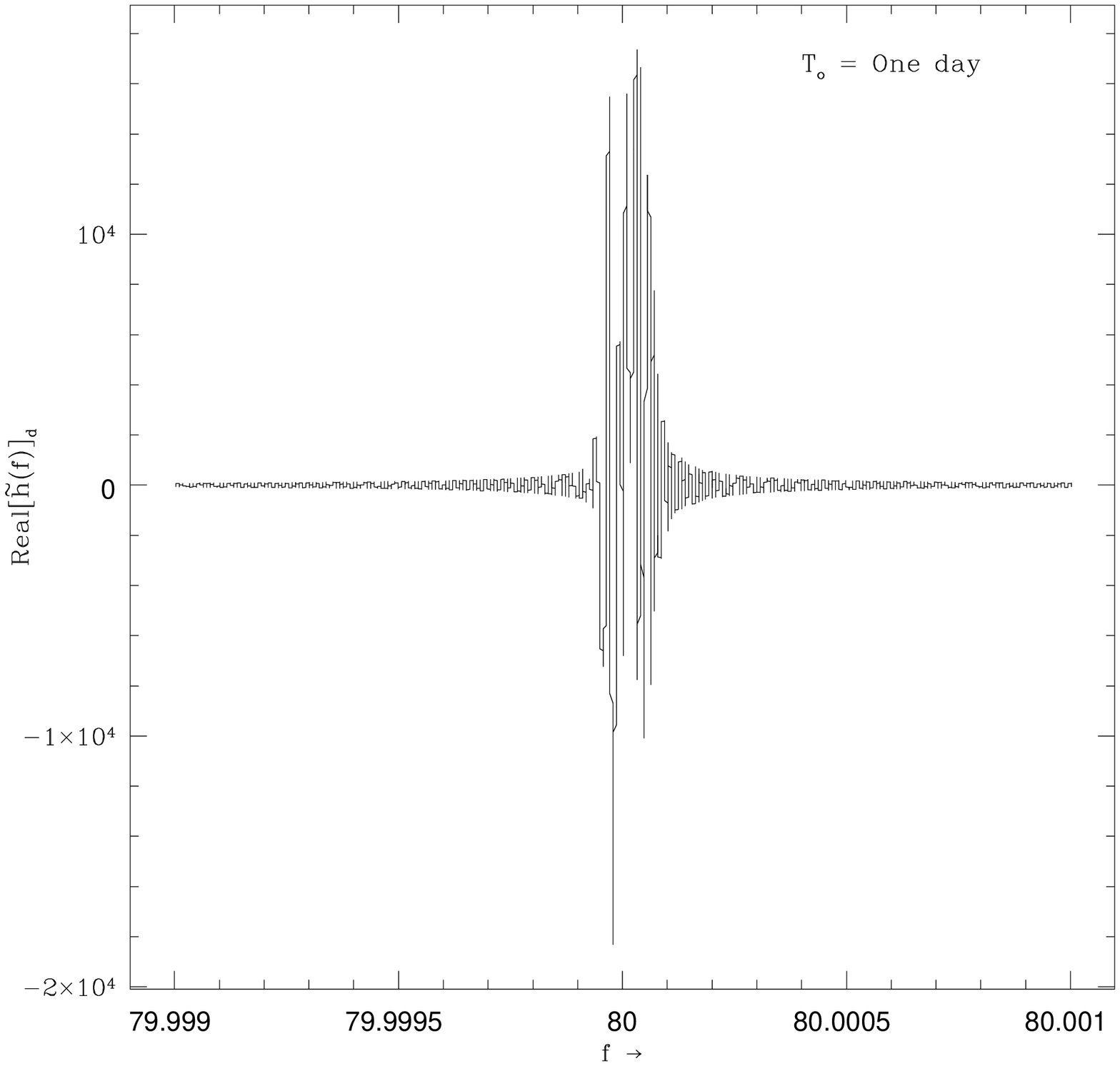,height=16.0cm}
\caption{FT of a FM signal frequency, $f_o = 80$ Hz, from a source located at 
$(\pi /36 , \pi )$ with a resolution of $10^{-7}$ Hz}
\label{fig:drealfmr7}
\end{figure}

\noindent The FT of the FM signal contains the
double series Bessel functions. The Bessel functions have contributions
due to the
rotational as well as the orbital motion of the earth.
It is remarked that
Jotania et al. (1996) have analysed FT of FM signal for one day observation
time. They have taken specific detector as well as source location. They have
also neglected the orbital motion. Our analysis generalizes their results.
We may now compute $\left[\tilde{h} (f)\right]_d$ and may plot its behaviour.
To achieve this we have made use of {\it Mathematica\/} software.
We know that the value of Bessel function decreases rapidly as its order
exceed the argument. Accordingly, one computes in practice only a finite
number of terms in the above infinite series. Figure~(\ref{fig:drealfmrt})
represents such a plot for

\begin{equation}
\label{eq:dayloc}
\left.\begin{array}{lll}
\vspace{0.2cm}
f_o = 80\; Hz\, , & h_{o_+} = h_{o_\times} = 1 & \\
\vspace{0.2cm}
\alpha = \pi /3\, , & \beta_o = \pi /4\, , & \gamma =  2\pi/5 \, , \\
\theta = \pi /36\, , & \phi = \pi\, , &  \psi = \pi /6. \\
\end{array} \right\}
\end{equation}

\noindent with a resolution equal to $1/T_o = 1.16 \times 10^{-5}$ Hz.
For the present case we found it sufficient to evaluate the
infinite series
for $k = - 21900 $ to $ + 21900$ and $ m = - 10$ to $+ 10$. Figures~(\ref{fig:drealfmr6}) and~(\ref{fig:drealfmr7}) represent the plot
$\;$of$\,$ the$\;$ FT$\;$ at resolution $10^{-6}$ Hz and $10^{-7}$ Hz. A careful
look at these plots reveals that the resolution of Fig.~(\ref{fig:drealfmrt}) 
does not represent the details of the 
dominant peaks around $f_o$, whereas, Fig.~(\ref{fig:drealfmr7}) does not
give any new behaviour as compared to Fig.~(\ref{fig:drealfmr6}). Hence,
we may say that a resolution of about $10^{-6}$ Hz is required to understand the correct
behaviour of the FT for one day observation data. In this reference
let us recall that the data analysis for Fast Fourier Transform  
(FFT) limits the resolution to $1/T_o$. However, the detector output may provide
us higher
resolution. Thus the semi-analytical analysis presented here may provide
more information as compared to FFT.

\section{Fourier transform of the complete response}
\label{sec:crd}
\indent The complete response $R(t)$, in view of
Eqs.~(\ref{eq:rt}),~(\ref{eq:fps}),~(\ref{eq:fcs}),~(\ref{eq:hpt})
and~(\ref{eq:hct}) may be written as

\begin{equation}
R(t) = R_+(t) + R_\times (t)\; ;
\end{equation}
\begin{eqnarray}
\label{eq:rpt}
R_+(t)& =& h_{o_+}\left[ F_{1_+}\cos 2\beta + F_{2_+}\sin 2\beta + F_{3_+}\cos \beta +
 F_{4_+}\sin \beta + F_{5_+} \right]\cos [\Phi (t)]\, , \\
 R_\times (t)& = & h_{o_\times}\left[ F_{1_\times}\cos 2\beta + F_{2_\times}\sin 2\beta +
F_{3_\times}\cos \beta + 
 F_{4_\times}\sin \beta + F_{5_\times}\right]\sin [\Phi (t)]
\label{eq:rct}
\end{eqnarray}

\noindent Here, for simplicity, we have taken the angle between the arms of
the detector equal to $\pi /2$ i.e. $\Omega = \pi /4$. Now the FT of the
complete response may be expressed as

\begin{eqnarray} \tilde {R}(f) &= &\tilde{R}_+(f) + \tilde{R}_\times (f)
\end{eqnarray}

\noindent Substituting $\beta$ as given
by~(\ref{eq:beta}) one obtains

\begin{eqnarray}
R_+(t)& =& \frac{h_{o_+}}{2}\left[e^{-i 2 \beta_o}( F_{1_+} +
i F_{2_+} ) e^{-i 2 w_{rot}t} + e^{i 2 \beta_o}( F_{1_+} - i F_{2_+} ) e^{i 2 w_{rot}t}
+\right. \nonumber \\
&&  e^{-i \beta_o}( F_{3_+} + i F_{4_+} )e^{-i w_{rot}t} + 
 e^{i \beta_o} ( F_{3_+} - i F_{4_+}) e^{i w_{rot}t}\frac{}{}
 + \left. 2 F_{5_+}\right]\cos [\Phi (t)]
\end{eqnarray}

\noindent and similar expression for $R_\times (t)$. Now it is straight-forward
to obtain the expressions for $\tilde{R}_+(f)$ and $\tilde{R}_\times (f)$ as

\begin{eqnarray}
\label{eq:rfpd}
\left[\tilde{R}_+(f)\right]_d & = & \frac{h_{o_+}}{2}\left[e^{-i 2 \beta_o}( F_{1_+} +
i F_{2_+} ) \left[\tilde{h} ( f + 2 f_{rot})\right]_d +  
 e^{i 2 \beta_o}( F_{1_+} - i F_{2_+} )\left[\tilde{h} ( f - 2 f_{rot})\right]_d\right.
 \nonumber \\
&& +\, e^{-i \beta_o}( F_{3_+} + i F_{4_+} )\left[\tilde{h} ( f +  f_{rot})\right]_d + 
  e^{i \beta_o} ( F_{3_+} - i F_{4_+}) \left[\tilde{h} ( f -  f_{rot})\right]_d \nonumber \\
&& \left. +\, 2 F_{5_+}\left[\tilde{h}(f)\right]_d\right]\, ;
\end{eqnarray}

 \begin{eqnarray}
\label{eq:rfcd}
\left[\tilde{R}_\times(f)\right]_d & = & \frac{h_{o_\times}}{2}\left[e^{-i 2 \beta_o}(
F_{2_\times} -i F_{1_\times} )\left[\tilde{h}( f + 2 f_{rot})\right]_d \right. 
- e^{i 2 \beta_o}( F_{2_\times} + i F_{1_\times} )\left[\tilde{h}( f -
2 f_{rot})\right]_d  \nonumber \\
&& +\, e^{-i \beta_o}( F_{4_\times} - i F_{3_\times} )\left[\tilde{h}( f +  f_{rot}) \right]_d 
 - e^{i \beta_o} ( F_{4_\times} + i F_{3_\times}) \left[\tilde{h}
( f -  f_{rot})\right]_d \nonumber \\
&& \left. - \, i 2 F_{5_\times}\left[\tilde{h}(f)\right]_d\right]
\end{eqnarray}

\noindent Collecting  our results the FT of the complete response
of the detector for one day time integration will be 

\begin{eqnarray}
\label{eq:rfd}
\left[\tilde{R}(f)\right]_d & = & \frac{1}{2}\left[e^{-i 2 \beta_o}\left[\tilde{h}( f
+ 2 f_{rot})\right]_d\left[ h_{o_+}( F_{1_+} + i F_{2_+} )
+ h_{o_\times} ( F_{2_\times} - i F_{1_\times} )\right]\right. +\nonumber \\
&& e^{i2\beta_o}\left[\tilde{h}( f - 2 f_{rot})\right]_d\left[ h_{o_+}
( F_{1_+} - i F_{2_+} )
- h_{o_\times} ( F_{2_\times} + i F_{1_\times} )\right] +\nonumber \\
& & e^{-i\beta_o}\left[\tilde{h}( f + f_{rot})\right]_d\left[ h_{o_+}( F_{3_+} + i F_{4_+} )
+ h_{o_\times} ( F_{4_\times} - i F_{3_\times} )\right] +\nonumber \\
&& e^{i\beta_o}\left[\tilde{h}( f - f_{rot})\right]_d\left[ h_{o_+}( F_{3_+} - i F_{4_+} )
- h_{o_\times} ( F_{4_\times} + i F_{3_\times} )\right] +\nonumber \\
&&\left.2 \left[ \tilde{h}(f)\right]_d\left[ h_{o_+}F_{5_+} - i h_{o_\times}F_{5_\times}\right]\right]
\end{eqnarray}

\noindent This shows that due to AM every Doppler modulated FM signal
will split in four additional lines at $f \pm 2 f_{rot}$ and $f \pm f_{rot}$, 
where $f_{rot}$ is the rotational
frequency of Earth ($f_{rot} \approx 1.16 \times 10^{-5}$ Hz).\\

\par
\begin{figure}
\epsfig{file=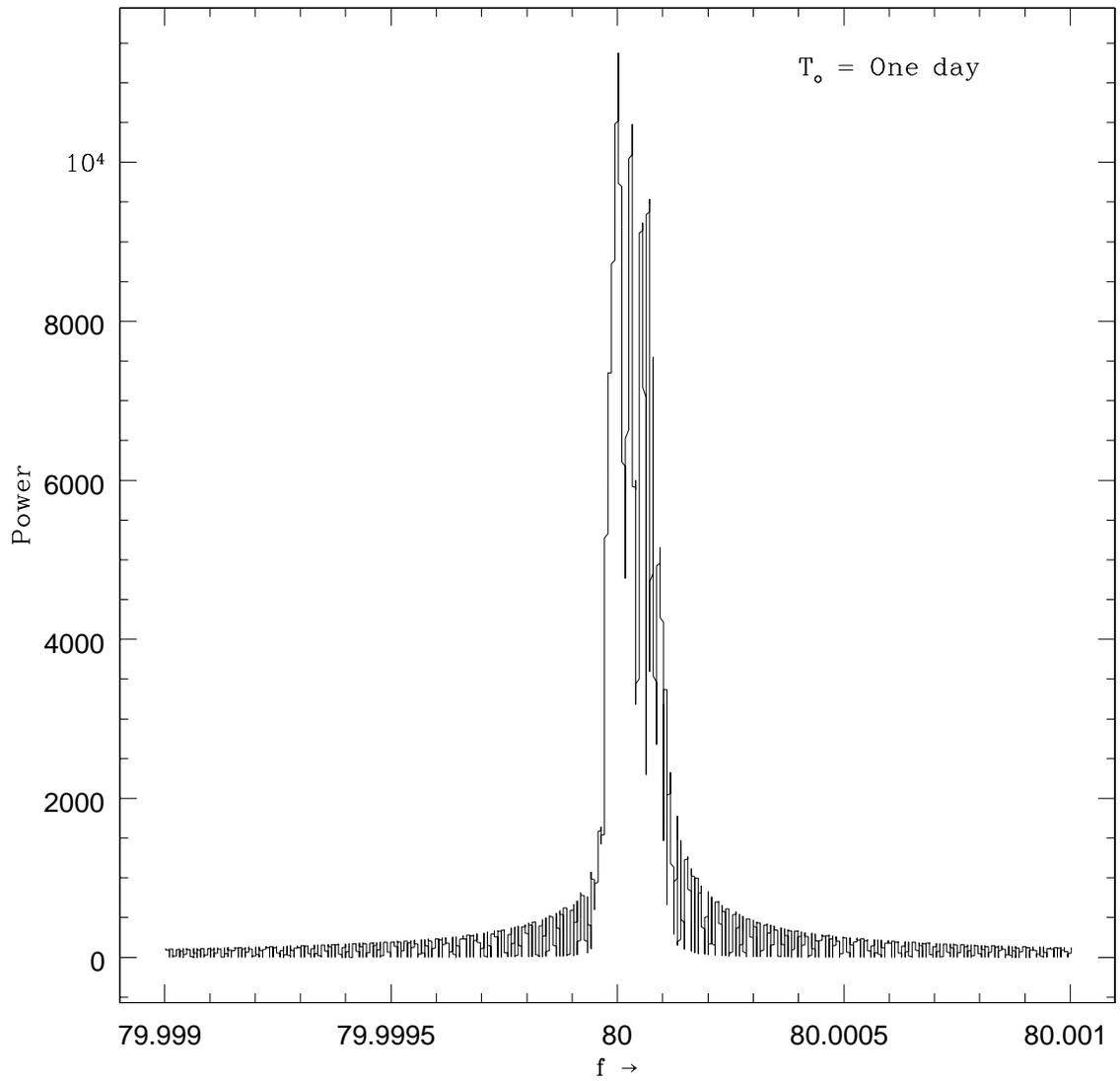,height=16.0cm}
\caption{Power spectrum of the complete response of a Doppler modulated signal frequency, $f_o = 80$ Hz, from 
a source located at $(\pi /36 , \pi )$ with a resolution of $10^{-7}$ Hz}
\label{fig:crd}
\end{figure}

\begin{figure}
\epsfig{file=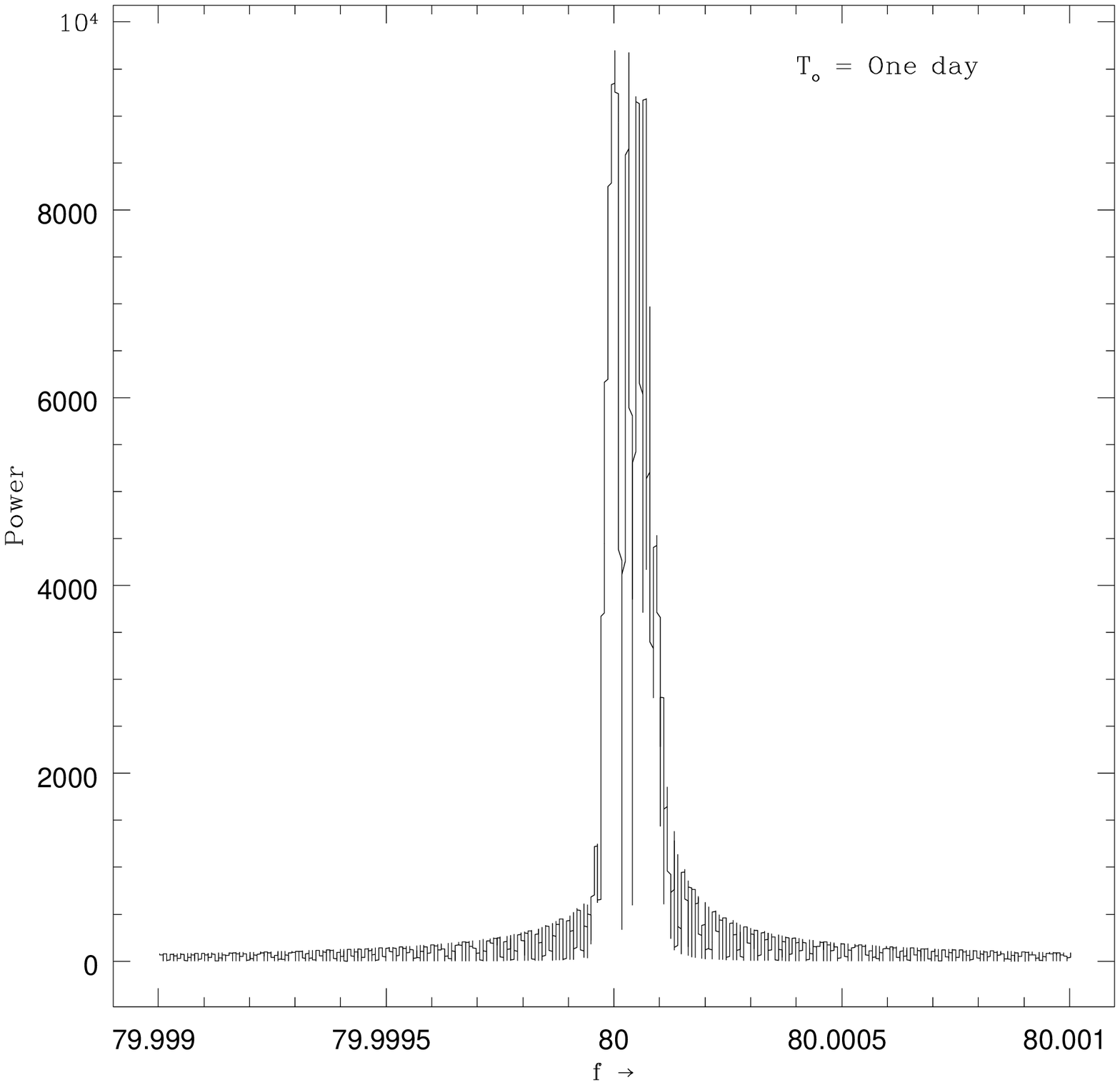,height=16.0cm}
\caption{Power spectrum of a Doppler modulated signal at frequencies  $f + 2f_{rot}$
of signal frequency, $f_o = 80$ Hz, from a source located at $(\pi /36 , \pi )$ with a resolution of $10^{-7}$ Hz}
\label{fig:crp2frd}
\end{figure}

\begin{figure}
\epsfig{file=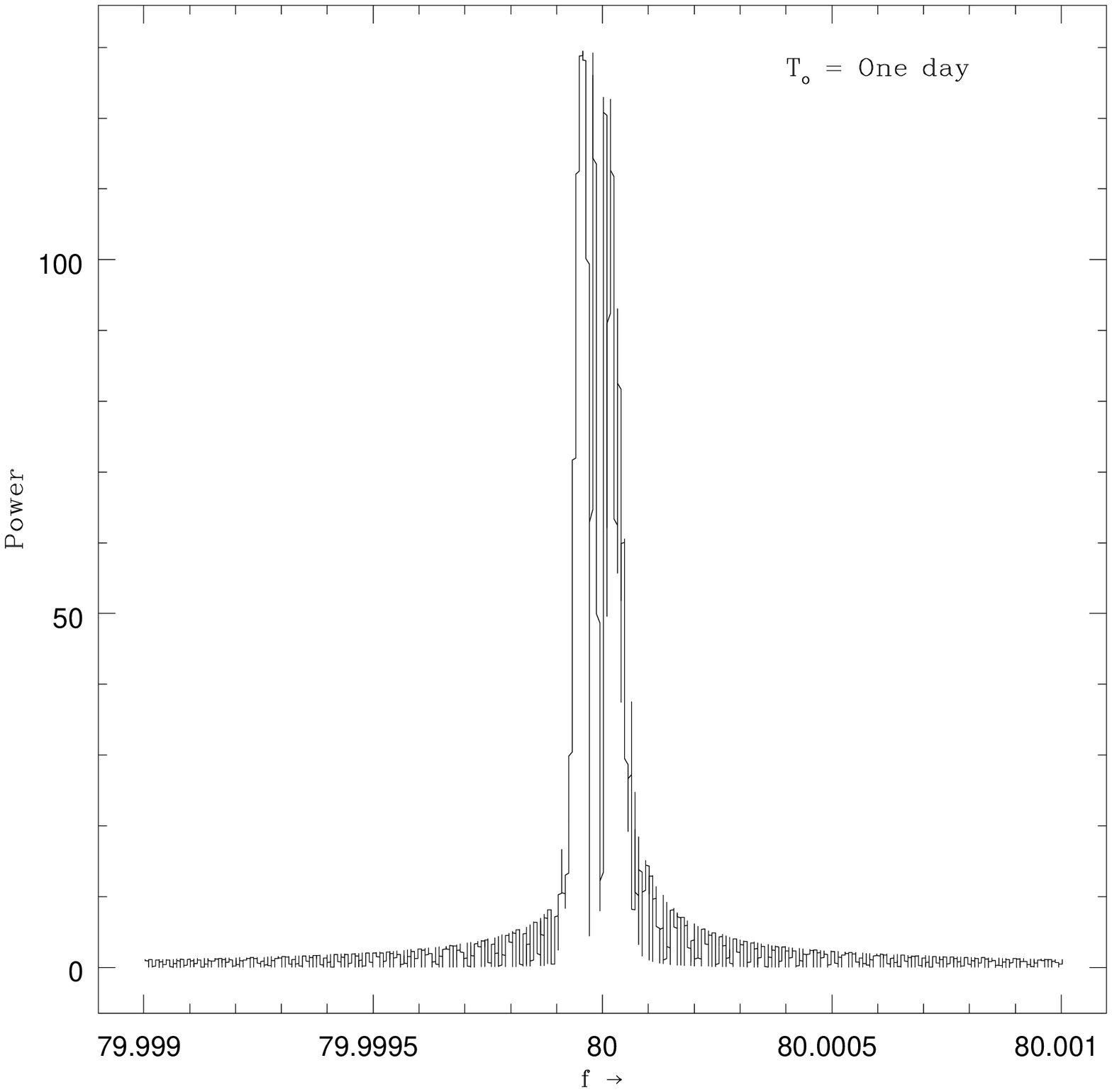,height=16.0cm}
\caption{Power spectrum of a Doppler modulated signal at frequencies  $f - 2f_{rot}$
of signal frequency, $f_o = 80$ Hz, from a source located at $(\pi /36 , \pi )$ with a resolution of $10^{-7}$ Hz}
\label{fig:crm2frd}
\end{figure}

\begin{figure}
\epsfig{file=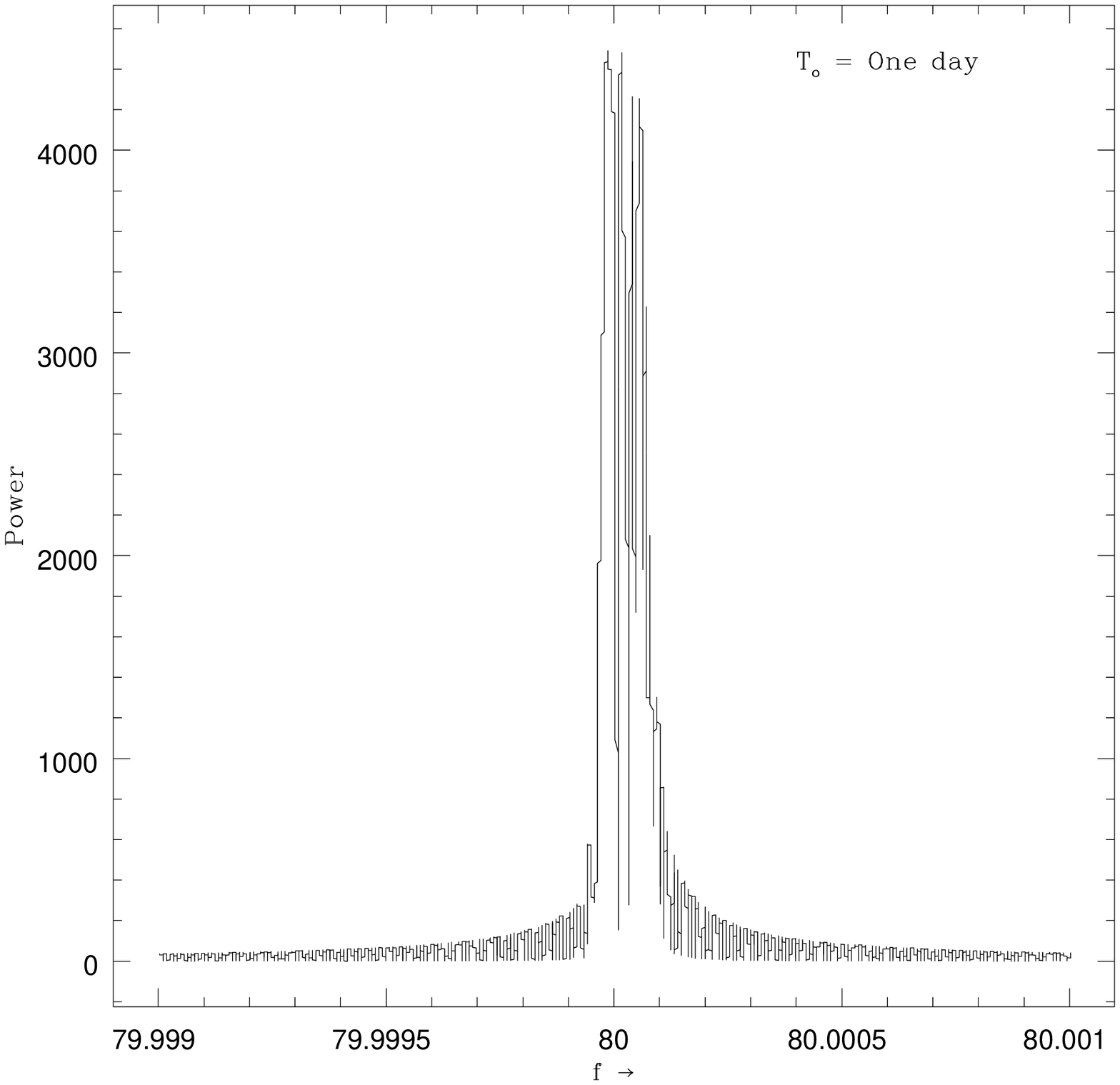,height=16.0cm}
\caption{Power spectrum of a Doppler modulated signal at frequencies  $f + f_{rot}$
of signal frequency, $f_o = 80$ Hz, from a source located at $(\pi /36 , \pi )$ with a resolution of $10^{-7}$ Hz}
\label{fig:crpfrd}
\end{figure}

\begin{figure}
\epsfig{file=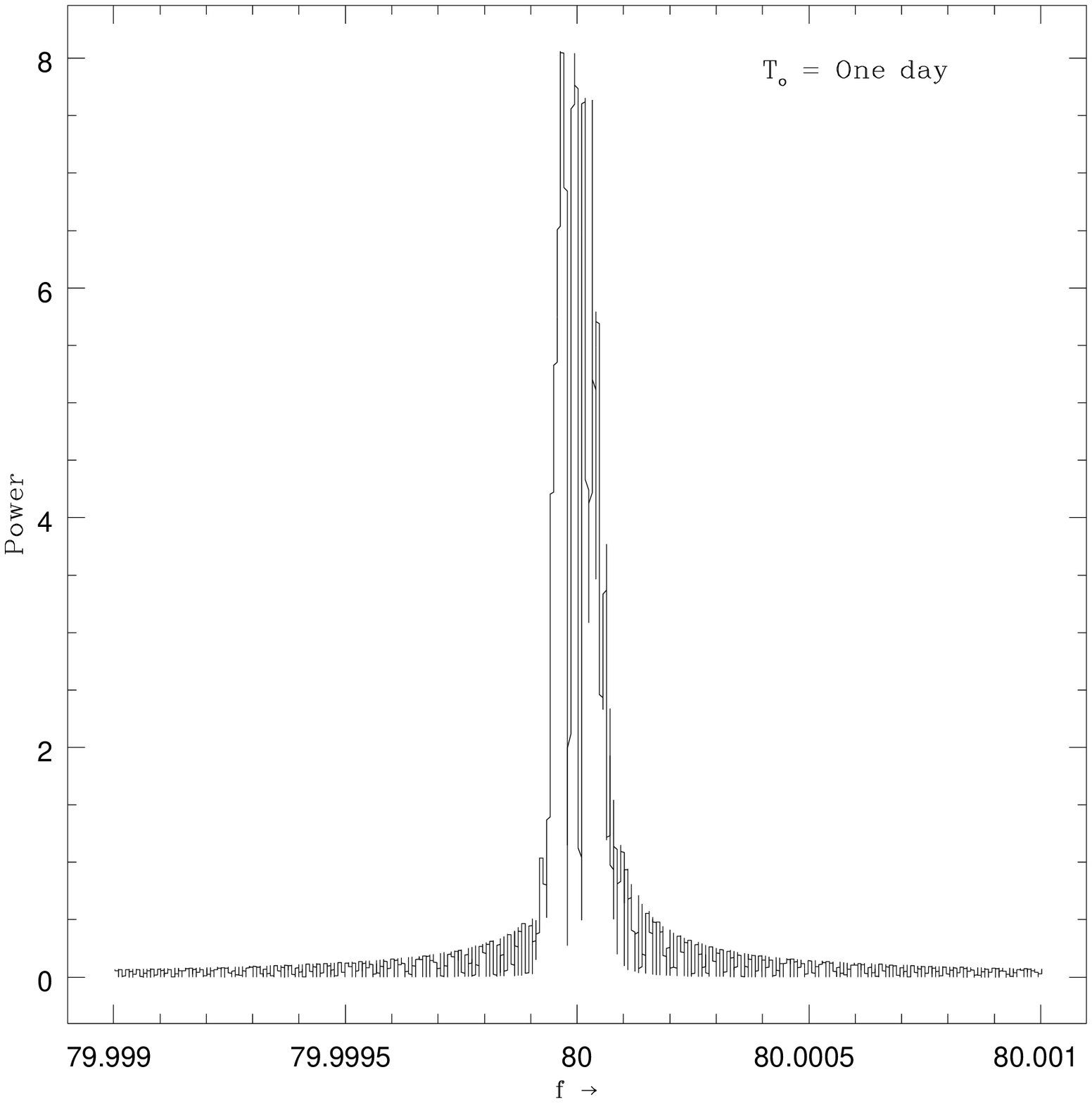,height=16.0cm}
\caption{Power spectrum of a Doppler modulated signal at frequencies  $f - f_{rot}$
of signal frequency, $f_o = 80$ Hz, from a source located at $(\pi /36 , \pi )$ with a resolution of $10^{-7}$ Hz}
\label{fig:crmfrd}
\end{figure}

\begin{figure}
\epsfig{file=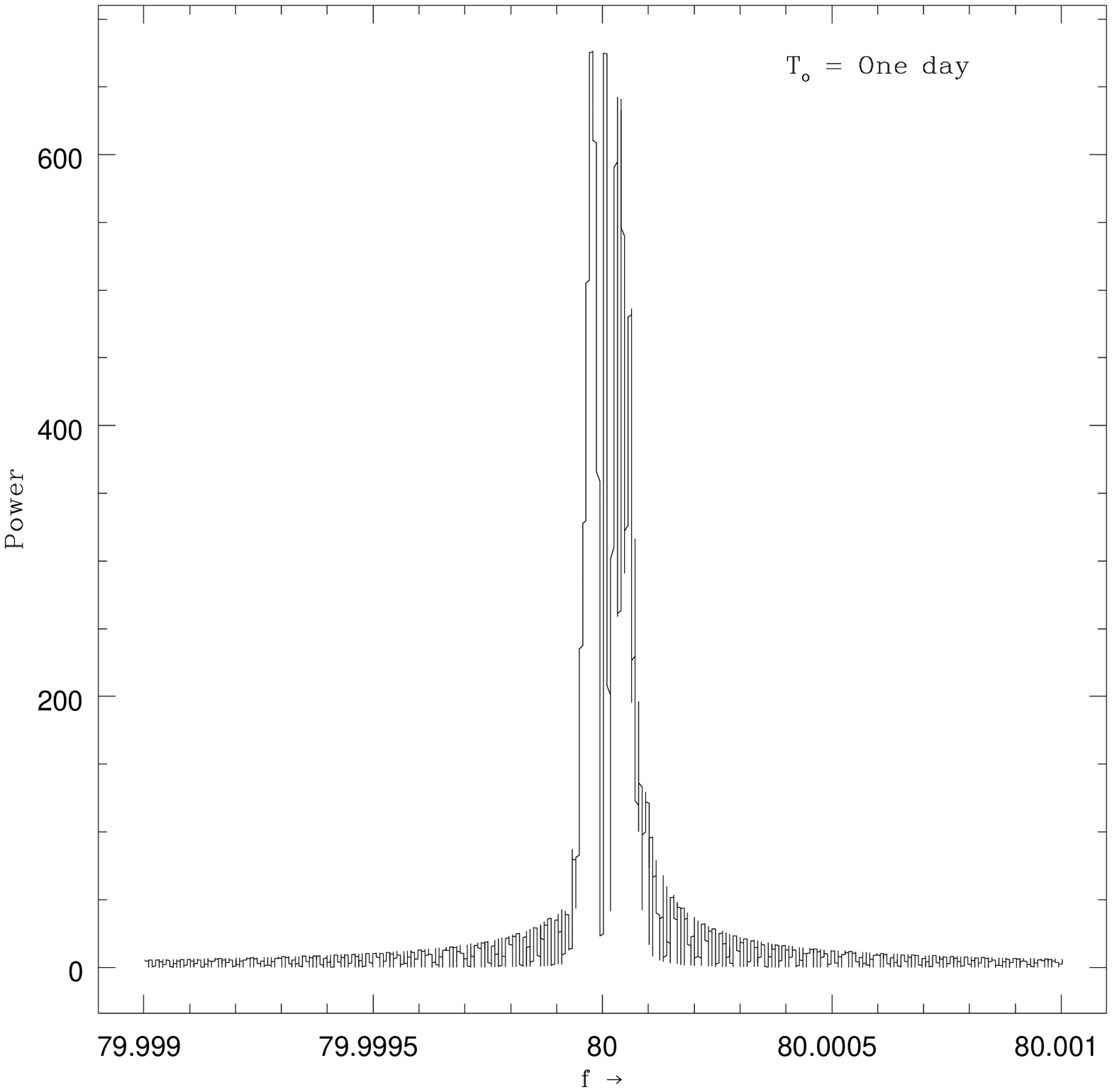,height=16.0cm}
\caption{Power spectrum of a Doppler modulated signal at frequencies  $f$ of signal frequency, $f_o = 80$ Hz, from a source
located at $(\pi /36 , \pi )$ with a resolution of $10^{-7}$ Hz}
\label{fig:cr0shiftd}
\end{figure}

 We have plotted in Fig.~(\ref{fig:crd}) the power spectrum of the noise 
free complete response of the signal for its various parameters as given
by~(\ref{eq:dayloc}). The contribution in the power spectrum of the modulation
at frequencies $f + 2f_{rot}$, $f - 2f_{rot}$, $f + f_{rot}$ and $f- f_{rot}$ and $f$ are
represented respectively in Figs.~(\ref{fig:crp2frd}),~(\ref{fig:crm2frd}),~(\ref{fig:crpfrd}),~(\ref{fig:crmfrd}) and~(\ref{fig:cr0shiftd}). It is observed that the most of the power 
will be at $f + 2f_{rot}$ and least power will be in $f - f_{rot}$.

\section{Conclusion}
\label{sec:sum3}
\indent In this paper we have considered the effect of earth's motion 
on the response of the detector through FT analysis. It can be
easily inferred from Eqs.~(\ref{eq:rfd}) and~(\ref{eq:hfpd},~\ref{eq:hfcd})
that the splitting of frequencies (i) in AM arises explicitly due to rotational motion and
(ii) in FM arises due to rotational as well as orbital motion of the earth.
In view of the fact that the data output at the detector is available in discrete
form, the analytical FT is not very convenient and one normally employs
the popular FFT. However, FFT has resolution limited 
to $1/T_o$. Further, it is important to understand for how much time one
can ignore the frequency shift arising due to Doppler effect. In fact, Schutz
(1991) has demonstrated that these effects due to rotational motion are
important after the time given by

\begin{equation}T_{max} = \left({\frac{2c}{\omega_{rot}^2 f_o R_e}}\right)^{1/2} \simeq
70\left( {\frac{f_o}{1 kHz}}\right)^{-1/2} min. \end{equation}

\noindent This means that for GW signal for frequency 80 Hz one has to take
into account these effects after data time $\simeq$ 4 hours. The
analytical FT studied in this paper leads to following inferences:
\begin{enumerate}[(i)]
\item FFT for one day observation data will not provide sufficient resolution
as to represent the correct picture of the frequency splitting.
\item The adequate resolution required for one day observation $\simeq 10^{-6}$ Hz.
\item The frequency split due to FM for frequency $f_o = 80$ Hz and source at
$(\theta , \phi )$ = $(\pi /36 ,\pi)$ is $\simeq$ 2 $\times 10^{-4}$ Hz and due
to AM is $\simeq$ $4.64 \times 10^{-5}$ Hz.
\item The drop in amplitude due to FM alone is about 56\%
\item The drop in amplitude due to AM alone is about 18\%.
\item The drop in amplitude for the complete response is about 74\%.
\item The maximum power due to AM is associated with $f_o + 2f_{rot}$.
\end{enumerate}

\noindent It is remarked that the drop of the amplitude in complete response is
severe both due to AM and FM as the relevant frequency range lies in the same region.\\

\par We have presented the FT analysis
assuming the phase of the GW to be zero at that instant $t = 0$. However, one
may relax this condition and may obtain the results easily by taking into
consideration the effects of change of the time origin.

\section*{Acknowledgments}
The authors are thankful to Prof. S. Dhurandhar, IUCAA, Pune and Dr. S.S.
Prasad, UNPG college, Padrauna for stimulating
discussions. The authors are extremely thankful to the anonymous referee 
for his hardwork in pinpointing the errors and making the detailed suggestions which 
resulted in the improvement of the paper. The authors are also thankful to IUCAA for providing hospitality
where major part of the work was carried out. This work is supported through
research scheme vide grant number SP/S2/0-15/93 by DST, New Delhi.

\begin{thebibliography}{99}
\bibitem{b1} Brady, P.R. and Creighton, 2000, \prd , 61, 082001. 
\bibitem{b2} Brady, P.R., et al., 1998, \prd , 57, 2101.
\bibitem{55} Goldstein, H., 1980, Classical Mechanics. Addison-wesley, New York.
\bibitem{b4} Jaranowski, P., et al, 1998, \prd , 58, 63001.
\bibitem{b5} Jaranowski, P. and Kr\'{o}lak, A., 1999, \prd , 59, 63003.
\bibitem{b6} Jaranowski, P. and Kr\'{o}lak, A., 2000, \prd , 61, 62001.
\bibitem{kn:b6a} Jotania, K., 1994, Some Aspects of Gravitational Waves
Signal Analysis from Coalescing Binaries and Pulsars, Ph.D. thesis, University of Pune, India,  
unpublished. 
\bibitem{b7} Jotania, K. and Dhurandhar, S.V., 1994, Bull. Astron. Soc. India, 22, 303.
\bibitem{b8} Jotania et al., 1996, Astronomy and Astrophysics, 306, 317-325.
\bibitem{b9} Kr\'olak, A., \pp 9903099 (1999).
\bibitem{b10} Schutz, B.F., and Tinto, M., 1987, MNRAS, 224, 131.
\bibitem{kn:102} Schutz, B.F., 1991, in The Detection of Gravitational Waves. ed. Blair, D.G.,
Cambridge University Press, Cambridge, England.
\end{thebibliography}
\end{document}